\documentclass[%
 reprint,
superscriptaddress,
 amsmath,amssymb,
 aps,
]{revtex4-2}

\usepackage{graphicx}
\usepackage{dcolumn}
\usepackage{bm}

\usepackage{xcolor}

\begin{document}

\preprint{APS/123-QED}

\title{Spontaneous breaking of time translation symmetry in a system without periodic external driving}

\author{T. T. Sergeev}
\affiliation{Moscow Institute of Physics and Technology, 141700, 9 Institutskiy pereulok, Moscow, Russia}
\affiliation{Dukhov Research Institute of Automatics (VNIIA), 127055, 22 Sushchevskaya, Moscow, Russia}
\affiliation{Institute for Theoretical and Applied Electromagnetics, 125412, 13 Izhorskaya, Moscow, Russia}
\affiliation{Institute of Spectroscopy Russian Academy of Sciences, 108840, 5 Fizicheskaya, Troitsk, Moscow, Russia}
\author{A. A. Zyablovsky}
\email{zyablovskiy@mail.ru}
\affiliation{Moscow Institute of Physics and Technology, 141700, 9 Institutskiy pereulok, Moscow, Russia}
\affiliation{Dukhov Research Institute of Automatics (VNIIA), 127055, 22 Sushchevskaya, Moscow, Russia}
\affiliation{Institute for Theoretical and Applied Electromagnetics, 125412, 13 Izhorskaya, Moscow, Russia}
\author{E. S. Andrianov}
\affiliation{Moscow Institute of Physics and Technology, 141700, 9 Institutskiy pereulok, Moscow, Russia}
\affiliation{Dukhov Research Institute of Automatics (VNIIA), 127055, 22 Sushchevskaya, Moscow, Russia}
\affiliation{Institute for Theoretical and Applied Electromagnetics, 125412, 13 Izhorskaya, Moscow, Russia}
\author{Yu. E. Lozovik}
\affiliation{Dukhov Research Institute of Automatics (VNIIA), 127055, 22 Sushchevskaya, Moscow, Russia}
\affiliation{Institute of Spectroscopy Russian Academy of Sciences, 108840, 5 Fizicheskaya, Troitsk, Moscow, Russia}
\affiliation{MIEM at National Research University Higher School of Economics, 123458, 34 Tallinskay, Moscow, Russia}

\date{\today}

\begin{abstract}
It is known that the spontaneous time translation symmetry breaking can occur in systems periodically driven at a certain period. We predict a spontaneous breaking of time translation symmetry in an atom-cavity system without external driving, in which a time scale is determined by the time of light bypass of the resonator. We demonstrate that there is a parameter range, in which a system state returns to its initial state only after two bypasses of the resonator. We believe that the predicted phenomenon opens a way to a new direction in the time crystal field.  
\end{abstract}

\maketitle

A concept of time crystal has been proposed by F. Wilczek \cite{wilczek2012}. Within this concept, a possibility of a spontaneous time translation symmetry breaking in quantum system is considered. In subsequent works \cite{bruno2013,nozieres2013,watanabe2015} the concept of the quantum time crystals has been criticized. In recent years, an idea of spontaneous time translation symmetry-breaking has been extended to Hamiltonian systems periodically driven at a certain period, $T_d$ ($\hat H\left( t \right) = \hat H\left( {t + {T_d}} \right)$) \cite{zhang2017,choi2017o,rovny2018,gong2018,zaletel2023}. In such systems, there exists a discrete time translation symmetry with respect to the time shift by $T_d$. Spontaneous time translation symmetry-breaking leads to the fact that some quantities in the system begin to change with a period that is a multiple of the driving period \cite{zaletel2023}. That is, in a parameter range, in which symmetry is broken, there is at least one quantity $Q\left( t \right)$ such that $Q\left( t \right) = Q\left( {t + m\,{T_d}} \right)$ where $m \geq 2$, is integer. In this regime, the time translation symmetry of the state is lower than that of the system \cite{zaletel2023}. Several intriguing applications of the systems with the spontaneous time translation symmetry-breaking have been proposed in recent years, for example, quantum sensing \cite{zhou2020,lyu2020} and quantum metrology \cite{preskill2018,choi2017}.

The time scale in the evolution of a system can arise not only under the influence of external driving, but also as a result of the interaction of the system with an environment of finite size. For example, the interaction of an excited two-level "atom" with a ring resonator (or cavity of finite length) leads to the transition of atom from the excited state to the ground one with the emission of a photon, which then propagates through the ring resonator and returns to the atom. A return time (a bypass time) is determined by the resonator size (${T_b} \propto L/c$) \cite{16}. After the bypass time, the emitted photon returns to the atom and the atom transits to the excited state. As a result, the dynamics of such systems can demonstrate periodical revivals and collapses with a period equal to the bypass time \cite{15,16,sergeev2022}. That is, the size of the resonator determines the time scale in the dynamics of the “atom”, $T_b$, and endows the system with a time translation symmetry.

In this letter, we demonstrate a possible of spontaneous breaking of time translation symmetry induced by the interaction with the environment of finite size. We consider system consisting of a two-level atom placed into a single mode cavity coupled with a ring-resonator. We demonstrate that in the considered structure, there exists a parameter range, at which photon emitted by the atom returns back, transitioning the atom to the excited state, after two bypasses of the resonator length. In such a parameter range, the atom evolution has symmetry only with respect to a shift by $2T_b$. This is an indication of time translation symmetry-breaking. We show that changing of the coupling strength between the atom and the cavity can lead to transition between regimes with broken and unbroken time translation symmetry. In one regime, the photon must bypass the ring resonator twice to return to its initial state. While in the other regime, one bypass of the resonator is sufficient to return to the initial state. We provide an explanation for the predicted phenomenon based on analysis of a dependence of the eigenstates on the interaction between atom, cavity mode and the ring resonator modes. We believe that the predicted phenomenon provides a new view on an emergence of spontaneous time translation symmetry-breaking in a system without periodic driving.

\begin{figure}[htbp]
\centering
\includegraphics[width=0.8\linewidth]{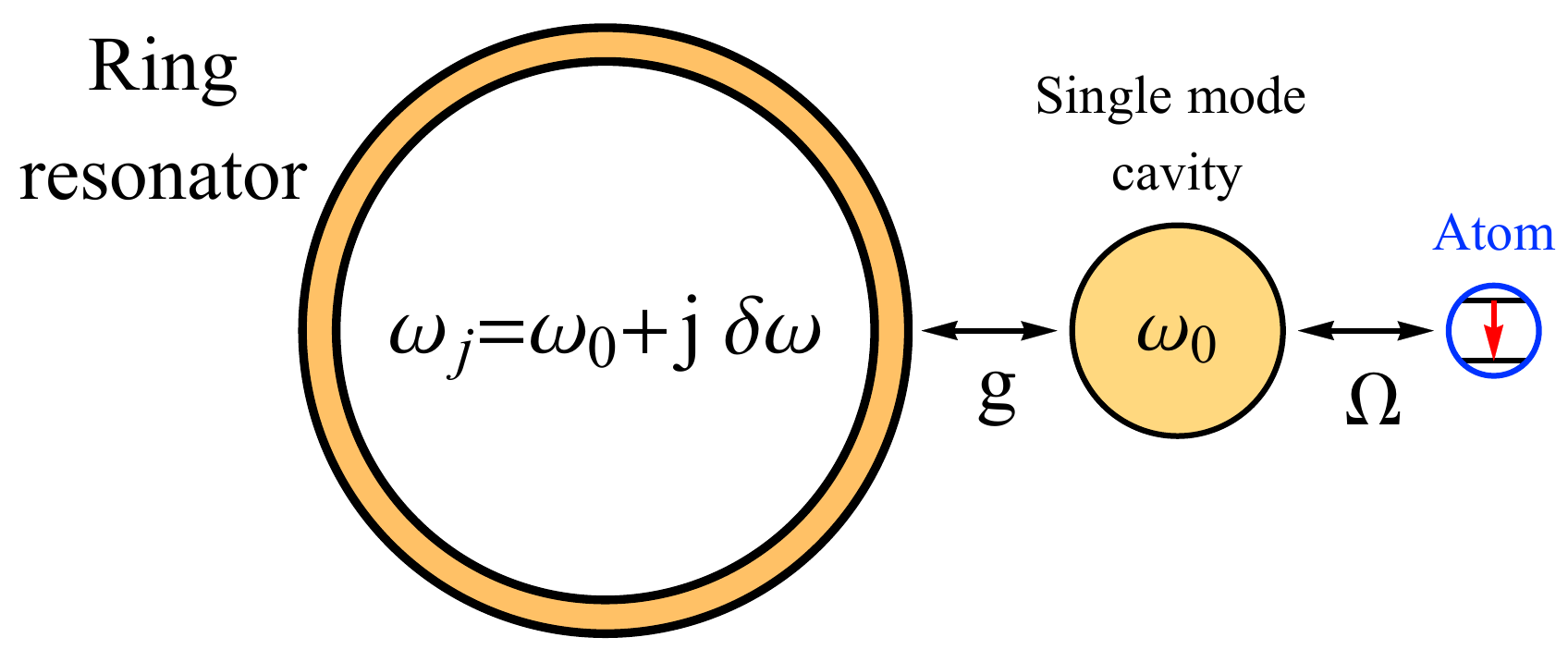}
\caption{Scheme of the system under consideration: a two-level atom placed in a single mode cavity coupled with a ring resonator.}
\label{fig:1}
\end{figure}

We consider a system consisting of a ring resonator with radius $R$, which couples with a single mode cavity. An active atom (two-level system) is placed into the cavity [Figure~\ref{fig:1}]. The role of atom can be played, e.g., by quantum dots, qubits operating in one quantum regime \cite{16}, etc. We consider that a transition frequency of the atom, $\omega_0$, coincides with the frequency of cavity mode. The frequencies of modes of the ring resonator are determined by the conditions $k\,L = 2\pi m,\,\,m = 1,2,...$, where $L = 2\pi R$ is the length of the ring resonator, $k = \omega \,n_{eff}/c$ is a wavenumber in the resonator and $n_{eff}$ is an effective refractive index. Specifically, we consider that the length of the ring resonator is selected so that the frequency of one of the modes coincides with the atomic frequency, $\omega_0$. That is, there is $m$ such that ${\omega _0} = m\,c/\left( {n_{eff}\,R} \right)$. Therefore, the frequencies of modes of the ring resonator can be given as ${\omega _j} = {\omega _0} + j\,\delta \omega $, where $\delta \omega  = c/\left( {n_{eff}\,R} \right)$ is a step between the mode's frequencies and $j$ is an integer number ($j = m - {\omega _0}n_{eff}R/c$).

To describe the system, we use the following Hamiltonian 
 \cite{14}
\begin{equation}
\begin{split}
\hat H = {\omega _{\,0}}\hat \sigma^\dag {\hat \sigma} + {\omega _{\,0}}\hat a^\dag {\hat a} + \Omega ({\hat a}{\hat \sigma}^\dag + \hat a^\dag \hat \sigma ) + \\
+ \sum\limits_{j = -N/2}^{{N/2}} {\omega _j \hat b_j^\dag {{\hat b}_j}} + \sum\limits_{j = -N/2}^{{N/2}} {{g_j}({\hat a}{\hat b}_j^\dag + \hat a^\dag \hat b_j )}
\end{split}
\label{eq:1}
\end{equation}
where ${\hat \sigma}$ and $\hat \sigma^\dag$ are the annihilation and creation operators for the two-level atom that satisfy the fermionic commutation relation, $\left\{ {\hat \sigma,{{\hat \sigma}^\dag }} \right\} = 1$. $\hat a$, ${\hat a^\dag }$  and $\hat b_j$, $\hat b_j^\dag$ are the annihilation and creation operators of the cavity mode and resonator modes, respectively, that satisfy the bosonic commutation relations, $\left[ {\hat a,{{\hat a}^\dag }} \right] = 1$ and $\left[ {\hat b_i,{{\hat b_j}^\dag }} \right] = \delta_{ij}$. $\Omega$ is the coupling strength between the atom and the cavity mode. We assume that every mode of the resonator interacts with cavity mode with the same coupling strength, i.e. $g_j = g$. We take into account $N$ modes of the ring resonator, the frequencies of which are closest to $\omega_0$ (i.e., $-N/2 \leq j \leq N/2$, $N >> 1$). To verify our results, we carried out calculations for various values of $N$. Our calculations show for sufficiently large $N$, the results cease to depend on $N$.

The evolution of the system (\ref{eq:1}) is described by the nonstationary Schr\"{o}dinger equation on the wave function $\vert \Psi(t) \rangle$. We assume that there is only one quantum in the system, which can be in the atom, in the cavity mode or in one of the resonator modes. Then we look for the wave function of the system in the following form \cite{14}:
\begin{equation}
\vert \Psi(t) \rangle = C_{\sigma} (t) \vert {e,0,0} \rangle + C_{a} (t) \vert {g,1,0} \rangle + \sum\limits_{j = -N/2}^{{N/2}} {C_j (t) \vert {g,0,1_j} \rangle}
\label{eq:2}
\end{equation}
where $\vert {e,0,0} \rangle$, $\vert {g,1,0} \rangle$ and $\vert {g,0,1_j} \rangle$ are states, in which an excitation quantum is in the atom, in the cavity mode and in one of the resonator modes, respectively. $C_{\sigma} (t)$, $C_a (t)$ and $C_j (t)$ are the probability amplitudes of finding an excitation quantum in the atom, in the cavity mode or in one of the resonator modes, respectively.

By substituting the wave function (\ref{eq:2}) into Schr\"{o}dinger equation we obtain the closed system of equations for  the probability amplitudes:
\begin{equation}
\frac{{d}}{{dt}} C_{\sigma} =  - i{\omega _{\,0}}{C_{\sigma}} - i\Omega C_a
\label{eq:3}
\end{equation}

\begin{equation}
\frac{{d}}{{dt}} C_a =  - i{\omega _{\,0}} C_a - i\Omega C_{\sigma} - \sum\limits_{j = -N/2}^{{N/2}} {i g C_j}
\label{eq:4}
\end{equation}

\begin{equation}
\frac{{d}}{{dt}} C_j =  - i{\omega _{\,j}} C_j - i g C_a
\label{eq:5}
\end{equation}
Hereinafter, we assume that at the initial moment of time the quantum of excitation is in the atom, i.e. $C_{\sigma} (0) = 1$ and $C_a (0) = C_{-N/2} (0) = ... = C_{N/2} (0) = 0$.

\begin{figure}[htbp]
\centering\includegraphics[width=\linewidth]{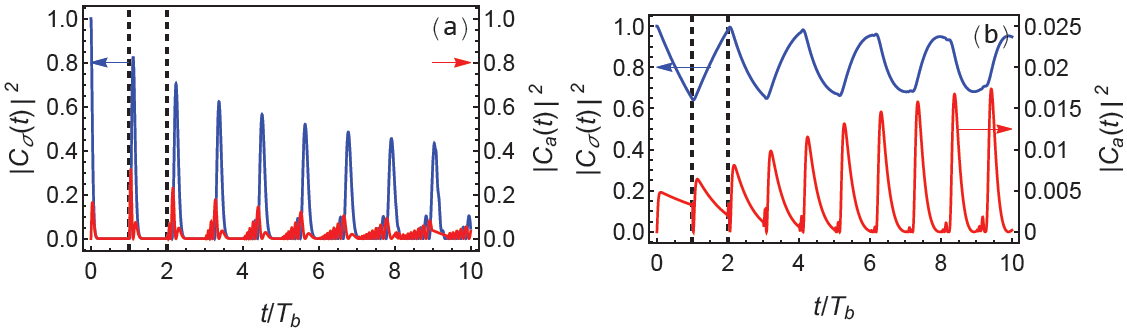}
\caption{The dependence of $\vert C_{\sigma} (t) \vert^2$ (blue line) and $\vert C_a (t) \vert^2$ (red line) on time at coupling strength (a) $\Omega = 5g >> g$ and (b) $\Omega=g/3 < g/\sqrt{2}$. Black dashed lines show the time of one and two bypass times. Here $N=100$, $g=0.003 \omega_0$, $\delta \omega = 0.002 \omega_0$, $T_b = 2 \pi/{\delta\omega} = 1000\pi {\omega_0}^{-1}$ and we take $\omega_0 = 1$.}
\label{fig:2}
\end{figure}

Since the system has discrete set of the frequencies, according to Poincar\'{e} recurrence theorem \cite{18}, there is Poincar\'{e} time in the system, when the system returns to a state equal to or close to the initial state. In the considered system, the return to initial state manifests itself in an appearance of collapses and revivals in the system dynamics [Figure~\ref{fig:2}]. Poincar\'{e} time usually can be estimated as ${T_P}=2\pi /\delta \omega $ \cite{15}, where $\delta\omega$ is a step between modes. This estimation coincides with the time of the ring resonator bypass $T_b$, which is determined as ${T_{b}} = n_{eff}\,L/c = 2\pi /\delta \omega $ (here we use the fact that $L = 2\pi R$ and $\delta \omega  = c/\left( {n_{eff}\,R} \right)$). That is, the size of the ring resonator determines a time scale of the considered system. Therefore, it can be expected that the dynamics of the system will approximately repeat with the step $T_b$. However, we will show that there is a parameter range, in which a doubling of Poincar\'{e} time is observed ($T_P \approx 2 T_b$). Such a doubling of Poincare time is a manifestation of time translation symmetry breaking.

If we look closer to the dynamics of the system (\ref{eq:3})-(\ref{eq:5}), we can see that the time dependence of $C_{\sigma} (t)$ and $C_a (t)$ significantly depends on the coupling strength $\Omega$ [see Figure~\ref{fig:2} (a) and (b)]. If the coupling strength $\Omega >> g$, the dynamics of the system, as expected, are ordinary oscillations with periodic revivals at times $T \approx m T_b = 2 \pi m/ \delta \omega$, $m \in \mathbb{N}$ [see Figure~\ref{fig:2}(a)]. However, if the coupling strength $\Omega < g/\sqrt{2}$, the dynamics of the system undergo a significant change. Namely, during the first bypass of electromagnetic wave, the probability $\left| C_\sigma \right|^2$ of the atom to be in the excited state slowly decreases. After the first bypass, $\left| C_\sigma \right|^2$ begins to increase and achieves unity after the second bypass. As a result, a doubling of Poincar\'{e} time can be seen in temporal dynamics of $\vert C_{\sigma} (t) \vert^2$ [see Figure~\ref{fig:2}(b)]. Although, the time of restoring of the state of the electromagnetic field remains the same [see Figure~\ref{fig:1}(b)]. When Poincar\'{e} time doubling occurs, the system loses its time translational symmetry, which is determined by the size of the ring resonator. Such a behavior can be associated with spontaneous breaking of time translation symmetry.

\begin{figure}[htbp]
\centering
\includegraphics[width=\linewidth]{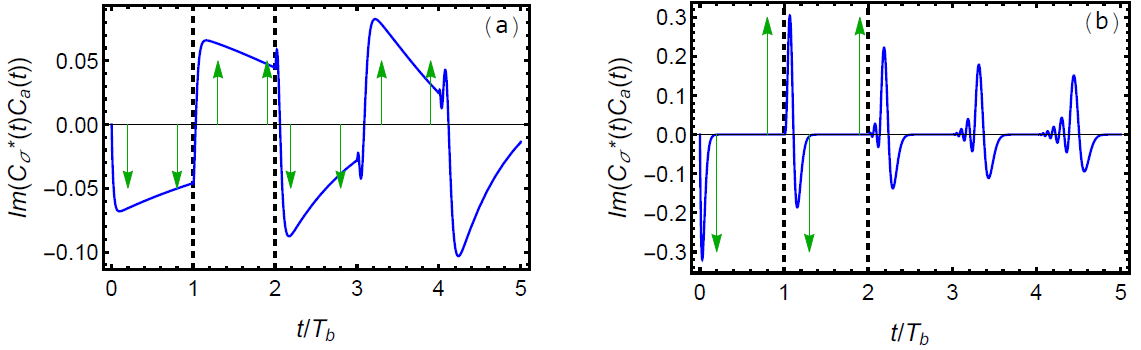}
\caption{The dependence of energy flow $\operatorname{Im}(\langle \hat \sigma^\dag \hat a \rangle)$ on time at coupling stregnth (a) $\Omega = g/3 < g/\sqrt{2}$ and (b) $\Omega = 5g >> g$. Green arrows show the direction of the energy flow between the cavity mode and the atom. Black dashed lines show the time of one and two bypass times. Here $N=100$, $g=0.003 \omega_0$, $\delta \omega = 0.002 \omega_0$, $T_b = 2 \pi/{\delta\omega} = 1000\pi {\omega_0}^{-1}$ and $\omega_0 = 1$.}
\label{fig:3}
\end{figure}

Spontaneous time translation symmetry breaking can be demonstrated more clearly by observing the energy flow between the atom and the cavity mode, $\operatorname{Im}  \left( \langle \hat{\sigma}^\dag \hat{a} \rangle \right)= \operatorname{Im} \left(C_{\sigma}^* (t) C_a (t) \right)$ \cite{41}. Our calculations show that during the system evolution, the energy flow changes the direction, which is determined by the phase difference between $C_a(t)$ and $C_{\sigma}(t)$. We calculate the dependence of $\operatorname{Im} \left(C_{\sigma}^* (t) C_a (t) \right)$ on the coupling strength $\Omega$ [see Figure~\ref{fig:3}]. It is seen [Figure~\ref{fig:3}(a)] that if the coupling strength $\Omega < g/\sqrt{2}$, then the direction of the energy flow comes to the initial state after a twice of the bypass time [see Figure~\ref{fig:3}(a)]. If the coupling strength $\Omega >> g$, then the energy flows in both sides during the one bypass [see Figure~\ref{fig:3}(b)].

The average direction of the energy flow during the bypasses can be characterized by the average phase difference between $C_a (t)$ and $C_{\sigma} (t)$ when averaging is carried over one bypass time $T_{b}$ and doubled bypass time $2 T_{b}$. If the system needs two bypass times to return to its initial state, then phase difference averaged over doubled bypass time should be equal to $0$. At the same time, while averaging is carried over one bypass time, there should be a non-zero phase difference. It means that system has not returned to its initial state and energy flow on average has preferred direction.

\begin{figure}[htbp]
\centering
\includegraphics[width=0.55\linewidth]{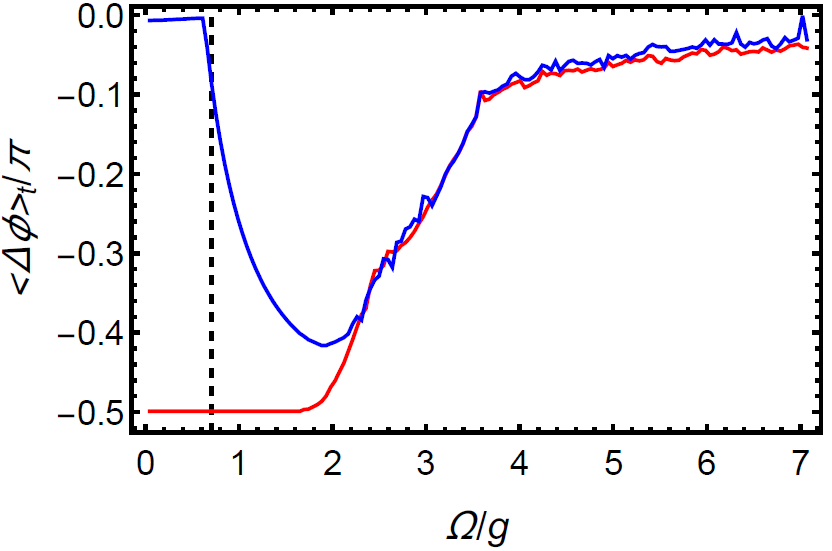}
\caption{The dependence of phase difference between $C_a (t)$ and $C_{\sigma} (t)$ averaged over the one bypass time (red line) and doubled bypass time (blue line), $\langle \Delta \phi {\rangle}_t = \langle \arg{(C_a (t) / C_{\sigma} (t))} {\rangle}_t$, on the coupling strength. Black dashed line shows the condition $\Omega = g/\sqrt{2}$. Here $N=100$, $g=0.003 \omega_0$, $\delta \omega = 0.002 \omega_0$ and $\omega_0 = 1$.}
\label{fig:4}
\end{figure}

Figure~\ref{fig:4} shows the dependence of the phase difference averaged over times $T_{b}$ and $2T_{b}$ on coupling strength. It is seen that there exists a clear transition when $\Omega$ is about $g/\sqrt{2}$. When $\Omega<g/\sqrt{2}$ the phase difference averaged over time $T_{b}$ does not equal zero. That is, the energy flow has a preferential direction during the one bypass. Although, during the two bypasses the preferential direction of the energy flow is absent. The increase of the coupling strength leads to disappearance of the preferential direction during the one bypass [see Figure~\ref{fig:4} for $\Omega>>g$]. The region of coupling strengths in which the phase difference averaged time $T_{b}$ is zero corresponds to an area with broken time symmetry.

The value of the coupling strength, $\Omega$, when the doubling of Poincar\'{e} time disappears, can be estimated from the following reasoning. 
From Figure~\ref{fig:2} it is seen that when Poincar\'{e} time is doubled, the amplitude, $C_{\sigma} (t)$, experiences a weak decay. Although when Poincar\'{e} time is not doubled ($\Omega >> g$) the amplitude decays to almost zero in the time intervals between revivals [Figure~\ref{fig:2}(a)]. That is, by the magnitude of the amplitude attenuation, it is possible to distinguish the region where doubling of Poincar\'{e} time is observed. Poincar\'{e} time doubling takes place when the coupling strength $\Omega \lesssim g$. In this case, the decay rate of amplitude $C_{\sigma} (t)$ at times $ t \leq T_{b}$ can be estimated as $\gamma' = 2 \Omega^2 / (g^2 T_{b})$ (i.e. $C_{\sigma} (t) \sim e^{-\gamma't}$, $t \leq T_{b}$) [see Appendix and \cite{17}]. Taking into account that the amplitude, $C_{\sigma} (t)$, experiences a weak decay, the condition for the disappearance of Poincar\'{e} time doubling can be estimated as the $\gamma' T_{b} \geq 1$, i.e. the amplitude $\sigma (t)$ experiences a weak decay at times $t \leq T_{b}$. The condition $\gamma' T_{b} = 1$ is satisfied when the coupling strength $\Omega = g/\sqrt{2}$. This condition is in good agreement with the behavior of the system dynamics [Figures~\ref{fig:2} and~\ref{fig:4}].

\begin{figure}[htbp]
\centering
\includegraphics[width=\linewidth]{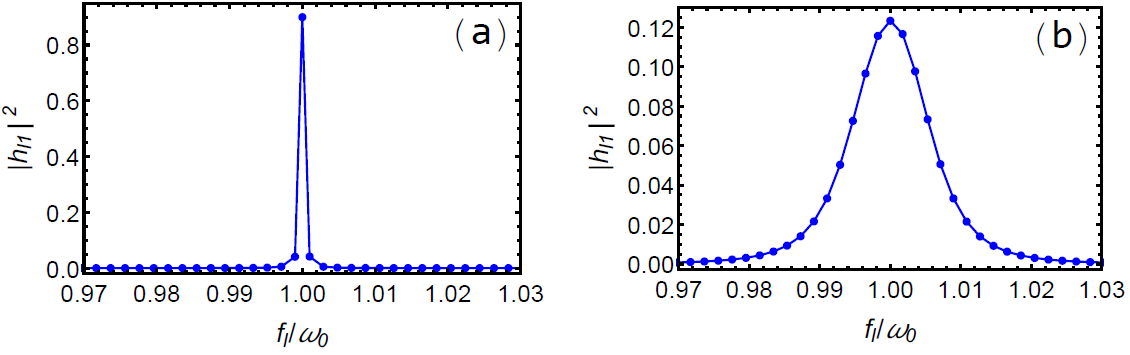}
\caption{The dependence of the squared amplitudes $\vert h_{l1} \vert^2$ of the first components of eigenvectors on eigenfrequencies $f_l$ at the coupling strength (a) $\Omega < g/\sqrt{2}$ and (b) $\Omega >> g$.}
\label{fig:5}
\end{figure}

The phenomenon of doubling of Poincar\'{e} time in the considered system can be explained by the structure of its eigenstates. We denote $\vec{h_l} = \left( h_{l1}, h_{l2}, h_{l3}, ..., h_{lN+3} \right)^T$ and ${\lambda}_l$ ($1\leq l \leq N+3$) as eigenvectors and eigenvalues of Eqns.~(\ref{eq:3})-(\ref{eq:5}); the components $h_{l1}$ determine contributions to the dynamics of $C_{\sigma} (t)$. Figure~\ref{fig:5} shows the distribution of the first components of the eigenvectors $\vert h_{l1} \vert^2$, $1 \leq l \leq N+3$ over the corresponding eigenfrequencies $f_l = -Im(\lambda_l)$ at coupling strengths $\Omega < g$ [Figure~\ref{fig:5} (a)] and $\Omega >> g$ [Figure ~\ref{fig:5} (b)].

At the coupling strength $\Omega < g$ components $\vert h_{l1} \vert^2$ form a very narrow peak at the frequency $\omega_0$. The linewidth of peak can be estimated as $\gamma'$. This means that the overwhelming contribution to the dynamics of the amplitude $C_{\sigma} (t)$ is given by modes with frequencies equal to or very close to the frequency $\omega_0$. Among the resonator modes, one mode with frequency $\omega_{j=0}=\omega_0$ satisfies this condition. At the same time, the interaction between atom with transition frequency $\omega_0$ and the cavity mode with frequency $\omega_0$ leads to appearance of additional eigenmode. This causes a doubling of the density of states near the frequency $\omega_0$. Note that the condition $\Omega \sim g/\sqrt{2}$ or $\gamma' T_{b} \sim 1$ is equivalent to the condition $\gamma'/{\delta\omega} \sim 1$, which corresponds to the case when number of modes at linewidth $\gamma'$ is about 1. Since Poincar\'{e} time is determined by the density of modes, it turns out that Poincar\'{e} time is doubled. In the case when the coupling constant $\Omega >> g$, this is incorrect, since other modes at frequencies different than $\omega_0$ also make a significant contribution to the dynamics, and the atom cannot be considered as an additional resonator mode. Therefore, there is no effective increase in the density of modes.

In conclusion, we predict a spontaneous breaking of time translation symmetry in an atom-cavity system, in which time scale arises from an interaction with a finite size of the reservoir, not from a periodical driving. We demonstrate that in the considered structure, there exists a parameter range, at which photon emitted by atom has to bypass the structure twice to return to its initial state. This means that a double bypass of the resonator is necessary to return to the initial state. Such a behavior is a manifestation of spontaneous breaking of time translation symmetry. We demonstrate that changing in the coupling strength between the atom and cavity mode leads to transition from the range, in which time translation symmetry is broken, to the one without breaking. Our results provide additional opportunities to study the spontaneous time translation symmetry-breaking in the system without periodic driving.

\section*{Acknowledgments}
The study was financially supported by a Grant from Russian Science Foundation (project No. 22-72-00026). T.T.S. and Yu.E.L. thank foundation for the advancement of theoretical physics and mathematics “Basis”.

\section*{Appendix}
We consider the closed system of equations~(\ref{eq:3})-(\ref{eq:5}) for the probability amplitudes of two-level system, cavity mode and modes of the ring resonator. In such a system, there are revivals of oscillations, which appear at times that are  multiples of Poincare time. Poincare time can be equal to one bypass time (${T_P} \approx {T_b} = \frac{{2\pi }}{{\delta \omega }}$) \cite{15} or doubled bypass time. At times less than the bypass time ($t \leqslant {T_b}$) the energy flows from the atom-cavity system to the ring resonator and, therefore, amplitudes of two-level system and cavity mode decay. To describe the dynamics of the system at times $t \leqslant {T_b}$, we switch to the resonator with an infinite number of modes and an infinitesimal density of states (i.e. $N \to \infty $ and $\delta \omega /{\omega _0} \to 0$). The transition to such resonator means that the return time is much longer than any observation time and it is possible to observe only the initial exponential decay stage. This stage is characterized by the decay rate, which also describes the dynamics of finite system~(\ref{eq:3})-(\ref{eq:5}) at times $t \leqslant {T_b}$ \cite{17}.

To derive the decay rate and the equations describing the evolution of system at times $t \leqslant {T_b}$, we first formally integrate the third equation in system~(\ref{eq:3})-(\ref{eq:5}):

\begin{equation}
{C_j}(t) = {C_j}(0){e^{ - i{\omega _j}t}} - ig\int\limits_0^t {d\tau {C_a}(\tau ){e^{ - i{\omega _j}\left( {t - \tau } \right)}}}
\label{eq:1A}
\end{equation}
Substituting Eq.~(\ref{eq:1A}) to Eqns.~(\ref{eq:3})-(\ref{eq:5}), we obtain
\begin{equation}
\frac{d}{{dt}}{C_\sigma } =  - i{\omega _0}{C_\sigma } - i\Omega {C_a}
\label{eq:2A}
\end{equation}
\begin{equation}
\frac{d}{{dt}}{C_a} =  - i{\omega _0}{C_a} - i\Omega {C_\sigma } - \sum\limits_j {{g^2}\int\limits_0^t {d\tau {C_a}(\tau ){e^{ - i{\omega _j}(t - \tau )}}} }
\label{eq:3A}
\end{equation}
where we used the fact that ${C_j}(0) = 0$.

Switching to slow amplitudes ${\tilde C_a}(t) = {C_a}(t){e^{i{\omega _0}t}}$ and ${\tilde C_\sigma }(t) = {C_\sigma }(t){e^{i{\omega _0}t}}$ we obtain
\begin{equation}
\frac{d}{{dt}}{{\tilde C}_\sigma } =  - i\Omega {{\tilde C}_a}
\label{eq:4A}
\end{equation}
\begin{equation}
\frac{d}{{dt}}{{\tilde C}_a} =  - i\Omega {{\tilde C}_\sigma } - \sum\limits_j {{g^2}\int\limits_0^t {d\tau {{\tilde C}_a}(\tau ){e^{ - i\left( {{\omega _j} - {\omega _0}} \right)(t - \tau )}}} }
\label{eq:5A}
\end{equation}
Applying the Born-Markov approximation \cite{carmichael,gardiner} and using the Sokhotskii-Plemelj formula \cite{plemelj}, we can calculate the integral and obtain the equations with decay:
\begin{equation}
\frac{d}{{dt}}{{\tilde C}_\sigma } =  - i\Omega {{\tilde C}_a}
\label{eq:6A}
\end{equation}
\begin{equation}
\frac{d}{{dt}}{{\tilde C}_a} =  - \gamma {{\tilde C}_a} - i\Omega {{\tilde C}_\sigma }
\label{eq:7A}
\end{equation}
where $\gamma  = \sum\limits_j {\pi {g^2}\delta \left( {{\omega _j} - {\omega _0}} \right)}$, $\delta (\omega )$ is Dirac’s delta-function.

Switching back to original variables, we finally obtain
\begin{equation}
\frac{d}{{dt}}{C_\sigma } =  - i{\omega _0}{C_\sigma } - i\Omega {C_a}
\label{eq:8A}
\end{equation}
\begin{equation}
\frac{d}{{dt}}{C_a} = \left( { - i{\omega _0} - \gamma } \right){C_a} - i\Omega {C_\sigma }
\label{eq:9A}
\end{equation}
Using the fact that ${\omega _j} = {\omega _0} + j\delta \omega$, the decay rate $\gamma$ can be estimated as $\gamma  = \frac{{\pi {g^2}}}{{\delta \omega }}$.

\begin{figure}[htbp]
\centering
\includegraphics[width=\linewidth]{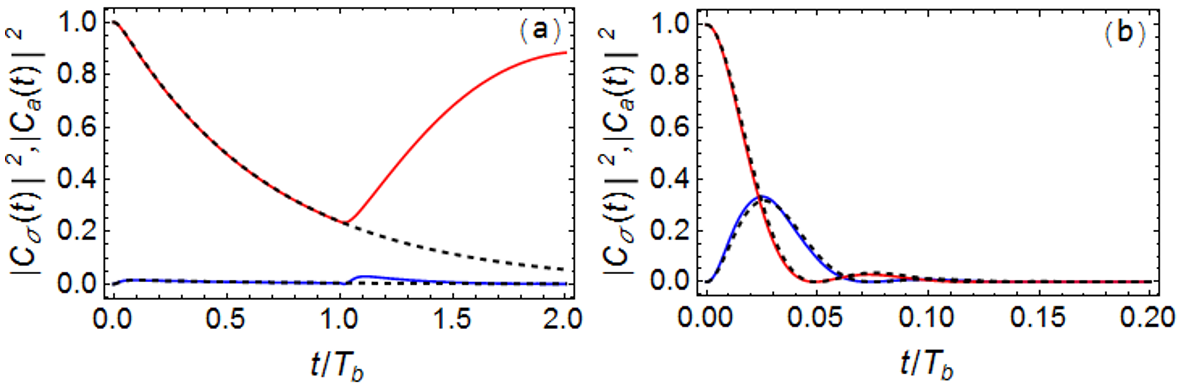}
\caption{The temporal dynamics of ${\left| {{C_\sigma }(t)} \right|^2}$ and ${\left| {{C_a}(t)} \right|^2}$ calculated by Eqns.~(\ref{eq:3})-(\ref{eq:5}) (solid red and blue lines, respectively) and Eqns.~(\ref{eq:8A}), (\ref{eq:9A}) (dashed black lines) at coupling strength $\Omega  < g/\sqrt 2$ (a) and $\Omega  = 5g >  > $ (b). Initial conditions ${C_\sigma }(0) = 1$, ${C_a}(0) = 0$ and ${C_j}(0) = 0$.}
\label{fig:1A}
\end{figure}

Therefore, Eqns.~(\ref{eq:8A}), (\ref{eq:9A}) with $\gamma  = \frac{{\pi {g^2}}}{{\delta \omega }}$ describe the dynamics of system~(\ref{eq:3})-(\ref{eq:5}) at times $t \leqslant {T_b}$. To demonstrate this, in [Figure~\ref{fig:1A} (a) and (b)] we present the temporal dynamics of ${\left| {{C_\sigma }(t)} \right|^2}$ and ${\left| {{C_a}(t)} \right|^2}$ calculated using Eqns.~(\ref{eq:3})-(\ref{eq:5}) and~(\ref{eq:8A}), (\ref{eq:9A}) at different coupling strengths $\Omega  < g/\sqrt 2$ (a) and $\Omega  = 5g >  > g$ (b).

To calculate the decay rate $\gamma '$ of two-level system at coupling strength $\Omega  < g/\sqrt 2$ we consider Eqns.~(\ref{eq:8A}), (\ref{eq:9A}) and calculate the corresponding eigenvalues and eigenvectors. The dynamics of system~(\ref{eq:8A}), (\ref{eq:9A}) can be described as follows:

\begin{equation}
\begin{gathered}
  \left( {\begin{array}{*{20}{c}}
  {{C_\sigma }(t)} \\ 
  {{C_a}(t)} 
\end{array}} \right) = {\alpha _ + }\left( {\begin{array}{*{20}{c}}
  {\frac{{i\gamma }}{{2\Omega }}\left( {1 - \sqrt {1 - \frac{{4{\Omega ^2}}}{{{\gamma ^2}}}} } \right)} \\ 
  1 
\end{array}} \right){e^{{\lambda _ + }t}} +  \hfill \\
  {\alpha _ - }\left( {\begin{array}{*{20}{c}}
  {\frac{{i\gamma }}{{2\Omega }}\left( {1 + \sqrt {1 - \frac{{4{\Omega ^2}}}{{{\gamma ^2}}}} } \right)} \\ 
  1 
\end{array}} \right){e^{{\lambda _ - }t}} \hfill \\ 
\end{gathered}
\label{eq:10A}
\end{equation}
where ${\lambda _ \pm } =  - i{\omega _0} - \frac{\gamma }{2}\left( {1 \pm \sqrt {1 - \frac{{4{\Omega ^2}}}{{{\gamma ^2}}}} } \right)$ eigenvalues of the system~(\ref{eq:8A}), (\ref{eq:9A}). The coefficients ${\alpha _ \pm }$ are determined by initial condition and using the condition ${\left( {\begin{array}{*{20}{c}}
  {{C_\sigma }(0)}&{{C_a}(0)} 
\end{array}} \right)^T} = {\left( {\begin{array}{*{20}{c}}
  1&0 
\end{array}} \right)^T}$ we obtain ${\alpha _ \pm } = \frac{{ \pm i\Omega /\gamma }}{{\sqrt {1 - \frac{{4{\Omega ^2}}}{{{\gamma ^2}}}} }}$.

Then for the amplitude of two-level system we get the following formula:
\begin{equation}
\begin{gathered}
  {C_\sigma }(t) = \frac{1}{2}\left( {1 - \frac{1}{{\sqrt {1 - \frac{{4{\Omega ^2}}}{{{\gamma ^2}}}} }}} \right){e^{{\lambda _ + }t}} \hfill \\
   + \frac{1}{2}\left( {1 + \frac{1}{{\sqrt {1 - \frac{{4{\Omega ^2}}}{{{\gamma ^2}}}} }}} \right){e^{{\lambda _ - }t}} \hfill \\ 
\end{gathered}
\label{eq:11A}
\end{equation}

At small values of coupling strength, i.e. when $\Omega  <  < \gamma$, we get ${C_\sigma }(t) \approx {e^{{\lambda _ - }t}}$. Then the decay rate of the amplitude of the two-level system is $\gamma ' =  - \operatorname{Re} {\lambda _ - } = \frac{\gamma }{2}\left( {1 - \sqrt {1 - \frac{{4{\Omega ^2}}}{{{\gamma ^2}}}} } \right) \approx {|_{\Omega  <  < \gamma }} \approx \frac{{{\Omega ^2}}}{\gamma }$. Using the fact that $\gamma  = \frac{{\pi {g^2}}}{{\delta \omega }} = \frac{{{g^2}{T_b}}}{2}$, we finally obtain $\gamma ' = \frac{{2{\Omega ^2}}}{{{g^2}{T_b}}}$.

\nocite{*}

\bibliography{apssamp}

\end{document}